\documentclass[prb,aps,twocolumn,draft,tightenlines,%
showpacs,floatfix,superscriptaddress]{revtex4}
\usepackage{psfig}
\usepackage{amssymb}

\begin{document}
\title{Many-body effect in spin dephasing in n-typed GaAs quantum
  wells}
\author{M. Q. Weng}%
\author{M. W. Wu}%
\thanks{Author to whom correspondence should be addressed}%
\email{mwwu@ustc.edu.cn}%
\affiliation{Structure Research Laboratory, University of Science \&%
Technology of China, Academia Sinica,  Hefei, Anhui, 230026, China}
\affiliation{Department of Physics, University of Science \&%
Technology of China, Hefei, Anhui, 230026, China}%
\altaffiliation{Mailing Address.}
\date{\today}
\begin{abstract}
  By constructing and numerically solving the kinetic Bloch equations
  we perform a many-body study of the spin dephasing due to the 
  D'yakonov-Perel' effect in $n$-typed GaAs (100) quantum wells for
  high temperatures. In our study, we include the
  spin-conserving scattering such as the electron-phonon, the
  electron-nonmagnetic impurity as well as  the electron-electron
  Coulomb scattering into consideration. The dephasing obtained from
  our theory 
  contains not only that due to the  effective spin-flip scattering
  first proposed by  D'yakonov and  Perel' [Zh. Eksp. Teor. Fiz. {\bf
    60}, 1954(1971)[Sov. Phys.-JETP {\bf 38}, 1053(1971)]], 
  but also the recently proposed many-body dephasing 
  due to the inhomogeneous broadening provided by the DP term
  [Wu, J. Supercond.:Incorp. Novel Mechanism {\bf 14}, 
  245 (2001); Wu and Ning, Eur. Phys. J. B {\bf 18}, 373 (2000)]. 
  Our result agrees with the experiment data [Malinowski {\em et al.},
  Phys. Rev. B {\bf 62}, 13034 (2000)] very well. We further show that
  in the case we study, the spin dephasing is dominated by the
  many-body effect.  
\end{abstract}

\maketitle

Spintronics,\cite{spintronics,wolf_sci_2001,spintronics_awsch} where
the spin degree of freedom of electrons is proposed to be used in
place of/in addition to the charge degree of freedom to add new
features and functionalities in semiconductor devices, has attracted
intensive attention recently. One of the important prerequisist for
the realization of the device application is to understand the spin
dephasing.  Experiments have shown long
spin lifetimes (up to $100$~ns) in $n$-typed 
semiconductors.\cite{damen,wagner,baumberg_1994_prl,baumberg_1994_prb,%
heberle,buss1,crooker_1996,crooker_1997,buss2,kikkawa1,kikkawa2,%
kikkawa3,ohno1,ohno,dzioev_2001,dzioev_2002} In theory, three
single-particle spin dephasing mechanisms, {\em ie.} the Elliot-Yafet
(EY) mechanism,\cite{yafet,elliot} the D'yakonov-Perel' (DP)
mechanism,\cite{dp} and the Bir-Aronov-Pikus (BAP)
mechanism\cite{bap}, have been proposed and discussed decades
ago.\cite{meier} All these spin dephasing mechanisms
require either spin-flip (SF) or effective SF scattering.  Recently,
Wu proposed a many-body spin dephasing mechanism where the spin
conservation (SC) scattering can also cause spin dephasing. The new
mechanism, which is caused by irreversibly disrupting the phases
between spin dipoles due to the inhomogeneous broadening together with
the SC scattering,\cite{wu_pss_2000,wu_ssc_2002,%
wu_jpsj_2001,wu_epjb_2000,wu_js_2001} has long been overlooked.  In
semiconductors, the inhomogeneous broadening can be introduced by the
energy dependence of the
$g$-factor\cite{wu_epjb_2000,wu_js_2001,bronold} and/or the momentum
${\bf k}$-dependence of the DP
term.\cite{wu_js_2001,wu_jpsj_2001,wu_ssc_2002} In spin
diffusion/transport, it may also come from the different spin
precession rates for different moment ${\bf k}$ along the spacial
gradience which we discussed closely in our recent
papers.\cite{weng_prb_2002,weng_jap_2003}

It is understood that for $n$-typed GaAs semiconductors, the DP
mechanism is the main spin dephasing mechanism at high
temperature. Very recently, we performed a systematic
investigation\cite{c0302330,c0210313,c0303169} of the spin dephasing
due to the DP effect in $n$-typed narrow quantum wells for high
temperatures ($\geq 120$~K) under magnetic fields in the Voigt
configuration by constructing and numerically solving the many-body
kinetic Bloch equations.\cite{wu_prb_2000,%
wu_pss_2000,wu_ssc_2002,wu_jpsj_2001,wu_epjb_2000,wu_js_2001} With all
of the SC scattering (electron-electron, electron-phonon and electron
nonmagnetic impurity scattering) included in our model, we are thus
able to study the spin dephasing under various conditions. The spin
dephasing time calculated from our theory contains not only the single
particle dephasing, but also the many-body dephasing. Moreover, from
this many-body theory, we are able to calculate the spin dephasing
with arbitrary initial spin polarization (up to 100\ \%) which cannot
be achieved from earlier single particle theory. Our results show many
dramastic many-body effects for both high and low spin
polarizations.\cite{c0302330,c0210313,c0303169} Many many-body effects
we predicted, especially those under high spin polarizations, have not
yet been measured experimentally.  In this communication, we will show
evidence of the many-body effect by comparing with the existing
experiments with small initial spin polarization.

We start our investigation from an $n$-typed (100) GaAs quantum well
(QW) with  
well width $a$. The growth direction is assumed to be $z$-axis. 
A moderate magnetic field {\bf B} is applied along the $x$ axis. 
Due to the confinement of the QW, the
momentum states along $z$ axis are quantized.  Therefore the electron
states are characterized by a subband index $n$ and a two dimensional
wave vector ${\bf k}=(k_x, k_y)$ together with a spin index $\sigma$.
For the QW with narrow width we study in the present paper, only the
lowest subband is populated and the transition
to the upper subbands is unimportant. With the DP term
included, the Hamiltonian of the electrons in the QW takes the form:
\begin{equation}
  H=\sum_{{\bf k}\sigma\sigma^{\prime}}\biggl\{
\varepsilon_{\bf k}+\bigl[g\mu_B{\bf B}+{\bf h}({\bf k})\bigr]
\cdot{\vec{\bf \sigma}_{\sigma\sigma^{\prime}}\over 2}\biggr\}
c^{\dagger}_{{\bf k}\sigma}c_{{\bf k}\sigma^{\prime}}+H_I.
\label{eq:hamiltonian}
\end{equation}
Here $\varepsilon_{{\bf k}}={\bf k}^2/2m^{\ast}$ is the energy of
electron with wavevector ${\bf k}$ and effective mass $m^{\ast}$.
$\vec{\bf \sigma}$ are the Pauli matrices. 
The DP term ${\bf h}({\bf k})$ 
can be written
as\cite{eppen,ivch}
\begin{eqnarray}
  &&h^{\mbox{BIA}}_x({\bf k})=\gamma k_x(k_y^2-\langle k_z^2\rangle), \;
  \nonumber\\
  &&h^{\mbox{BIA}}_y({\bf k})=\gamma k_y(\langle k_z^2\rangle-k_x^2), \;
\nonumber\\
&&  h^{\mbox{BIA}}_z({\bf k})=0\ .
  \label{eq:dp}
\end{eqnarray}
Here $\langle k^2_z\rangle$ represents the average of the operator
$-({\partial\over\partial z})^2$ over the electronic state of the
lowest subband and is therefore $(\pi/a)^2$.
$\gamma=(4/3)(m^{\ast}/m_{cv})(1/\sqrt{2m^{\ast
3}E_g})(\eta/\sqrt{1-\eta/3})$ and $\eta=\Delta/(E_g+\Delta)$, in
which $E_g$ denotes the band gap; $\Delta$ represents the spin-orbit
splitting of the valence band; 
and $m_{cv}$ is a constant close in magnitude to the free
electron mass $m_0$.\cite{aronov} 
The interaction Hamiltonian $H_I$ is composed of Coulomb interaction
$H_{ee}$, electron-phonon interaction $H_{ph}$, as well as
electron-impurity scattering $H_i$. Their expressions can be found in
textbooks.\cite{haug,mahan} 

We construct the kinetic Bloch equations by the nonequilibrium Green
function method\cite{haug} as follows:
\begin{equation}
  \label{eq:bloch}
  \dot{\rho}_{{\bf k},\sigma\sigma^{\prime}}
  =\dot{\rho}_{{\bf k},\sigma\sigma^{\prime}}|_{\mbox{coh}}
  +\dot{\rho}_{{\bf k},\sigma\sigma^{\prime}}|_{\mbox{scatt}}\ .
\end{equation}
Here $\rho_{{\bf k}}$ represents the single particle density
matrix. The diagonal elements describe the electron distribution
functions $\rho_{{\bf k},\sigma\sigma}=f_{{\bf k}\sigma}$. The
off-diagonal elements 
$\rho_{{\bf k},{1\over 2}-{1\over2}}\equiv\rho_{{\bf k}}$ 
describe the inter-spin-band polarizations
(coherence) of the spin coherence.\cite{wu_prb_2000} Note that 
$\rho_{{\bf k},-{1\over 2}{1\over 2}}\equiv 
\rho^{\ast}_{{\bf k},{1\over 2}-{1\over 2}}=\rho^{\ast}_{{\bf k}}$. 
Therefore, $f_{{\bf k}\pm{1\over 2}}$ and $\rho_{{\bf k}}$ are the 
quantities to be determined from Bloch equations. 

The coherent parts of the equation of motion for the electron
distribution function and the spin coherence are given by 
\begin{widetext}
\begin{eqnarray}
  \label{eq:f_coh}
 && {\partial f_{{\bf k},\sigma}\over \partial t}|_{\mbox{coh}}=
-2\sigma\bigl\{[g\mu_BB+h_x({\bf k})]\mbox{Im}\rho_{{\bf k}}+h_y({\bf k})
\mbox{Re}\rho_{{\bf k}}\bigr\}
+4\sigma\mbox{Im}\sum_{{\bf q}}V_{{\bf q}}\rho^{\ast}_{{\bf k}+{\bf
    q}} \rho_{{\bf k}},\\
 \label{eq:rho_coh}
&&  {\partial \rho_{{\bf k}}\over \partial t}\left |_{\mbox{coh}}\right. =
  {1\over 2}[ig\mu_B B + ih_x({\bf k}) + h_y({\bf k})]
  (f_{{\bf k}{1\over 2}}-f_{{\bf k}-{1\over 2}})
  +i\sum_{{\bf q}}V_{\bf q}\bigl[(f_{{\bf k}+{\bf q}{1\over 2}}
  -f_{{\bf k}+{\bf q}-{1\over 2}})\rho_{{\bf k}}
  -\rho_{{\bf k}+{\bf q}}(f_{{\bf k}{1\over 2}}
  -f_{{\bf k}-{1\over 2}})\bigr],
\end{eqnarray}
\end{widetext}
where $V_{{\bf q}}=4\pi e^2/[\kappa_0(q+q_0)]$ is the 2D Coulomb
matrix element under static screening.
$q_0=(e^2m^{\ast}/\kappa_0)\sum_{\sigma}f_{{\bf k}=0,\sigma}$ and
$\kappa_0$ is the static dielectric constant.  The first term on the
right hand side (RHS) of Eqs.~(\ref{eq:f_coh}) and (\ref{eq:rho_coh})
describes spin precession of electrons under the magnetic field ${\bf
B}$ as well as the effective magnetic field ${\bf h}({\bf k})$ due to
the DP effect. The rest terms are the contribution to the spin precession
from the HF term of the electron-electron interaction.
$\dot{\rho}_{{\bf k},\sigma\sigma^{\prime}}|_{\mbox{scatt}}$ stands for
the scattering term. In our theory, we include all of the SC
scattering, {\em i.e.}, the electron-non-magnetic impurity,
electron-phonon as well as the electron-electron scattering. Their
expressions are listed in detail in
Refs.~\onlinecite{c0302330,c0210313,c0303169}, and will not be
repeated here.
 
The initial conditions are taken at $t=0$ as: 
\begin{eqnarray}
  &&\rho_{\bf k}|_{\rm t=0} = 0
  \label{eq:rho_init}\\
&&  f_{{\bf k}\sigma}|_{\rm t=0} = 1/\bigl\{\exp[(\varepsilon_{\bf
    k}-\mu_{\sigma})/k_BT]+1\bigr\} 
  \label{eq:fk_init}
\end{eqnarray}
where $\mu_\sigma$ is the chemical potential for spin $\sigma$. The
condition $\mu_{\frac{1}{2}}\neq\mu_{-\frac{1}{2}}$ gives rise to the
imbalance of the electron densities of the two spin
bands. Eqs.~(\ref{eq:bloch}) through (\ref{eq:rho_coh}) together
with the initial conditions Eqs.~(\ref{eq:rho_init}) and
(\ref{eq:fk_init}) comprise the complete 
set of kinetic Bloch equations of our investigation.

The kinetic Bloch equations form a set of nonlinear equations. The DP
term in coherent part of the Bloch equation tends to drive the
electron distribution out of equilibrium Fermi distribution, while the
scattering term attempts to randomize electrons in ${\bf
k}$-space. Therefore, these two parts are coupled in the Bloch
equations and the equations have to be solved self-consistently.  This
self-consistent solution can be performed numerically through the
method that has been laid out in detail in the previous paper on the
DP mechanism in 3D systems.\cite{wu_pss_2000} By numerically solving
these coupled equations, we are thus able to study the spin procession
between spin-up and -down bands.  As discussed in the previous
papers,\cite{wu_prb_2000,kuhn} one can obtain the spin dephasing time
as the inverse of the slope of the envelope of the incoherently summed
spin coherence $\rho(t)=\sum_{{\bf k}}|\rho_{{\bf k}}|$. This spin
dephasing time consists both the single-particle
effective-SF-scattering-induced spin dephasing time and the many-body
spin dephasing time. We will compare our spin dephasing time with the
experiments as well as that predicted by the widely used single
particle theory:\cite{wu_jpsj_2001,averkiev,lau}
\begin{equation}
  \label{eq:taus_ani}
  {1\over \tau} = {\int_0^{\infty} d E_{k} \bigl( 
    f_{k{1\over 2}} - f_{{k}-{1\over 2}}\bigr)\Gamma(k)
    \over \int_0^{\infty} d E_{k} \bigl( f_{k{1\over
    2}}-f_{{k}-{1\over 2}}\bigr)}, 
\end{equation}
with
\begin{eqnarray}
  \Gamma(k)& =&  {2\tau_{1}(k)}\Bigl[\bigl(\gamma\langle
    k_z^2\rangle\bigr)^2 k^2 - 
    {1\over 2}\gamma\langle k_z^2\rangle k^4 \nonumber\\
&&\mbox{}\hspace{1cm} + {{1+\tau_3(k)/\tau_1(k)}\over 16}\gamma^2 k^6\Bigr]
  \label{eq:gammak}\\
  \label{eq:tau_n}
  \tau_n^{-1}(k)& =&
  \int_0^{2\pi}\sigma(E_k,\theta)[1-\cos(n\theta)]d\theta\ .
\end{eqnarray}
Here $\sigma(E_k,\theta)$ stands for the scattering cross-section. 

We compare the spin dephasing times calculated from our many-body
theory in two GaAs QW's for the well width $a$ being 6\ nm and 10\ nm
with those from the experiment by Malinowski {\em et al}.\cite{mali}
In the experiment the spin dephasing times for 15
and 20\ nm wells have also been measured. However, for these wider
width QW's, one needs more 
subbands than just the lowest one as we use in the present
paper.  The electron density is $N_e=2\times 10^{11}$~cm$^{-2}$ which
is the background excitation measured in the experiment for both
samples. The material parameters of GaAs for our calculation are
tabulated in Table~\ref{table1}.\cite{made} The impurity densities
$N_i$ are taken to be $0.1 N_e$ for $a=6$~nm QW and $0.05 N_e$ for
$a=10$~nm QW.  It is noted that here we use the elastic
impurity scattering to simulate also the electron scattering by the
interface roughness. Therefore it is understood that the impurity
density for the 6~nm QW is larger than that for the 10\ nm well as
narrower QW suffers more scattering from the interface roughness.

\begin{table}[htb]
  \centering
  \begin{tabular}{lllllll}
    \hline\hline
    $\kappa_\infty$ & \mbox{}\hspace{1.25cm} &
    10.8 & \mbox{}\hspace{1.25cm} &
    $\kappa_0$ & \mbox{}
    \hspace{1.25cm} & 12.9\\ 
    
    $\omega_0$ & & 35.4~meV & & $m^*$ & &0.067~$m_0$\\
    $\Delta$ & &0.341~eV & &$E_g$ & &1.55~eV\\
    $g$ & &0.44 &&$m_{cv}$& & 1.25 $m_0$\\
    \hline\hline
  \end{tabular}
  \caption{Parameters used in the numerical calculations}
  \label{table1}
\end{table}

\begin{figure}[htb]
  \centering
  \psfig{file=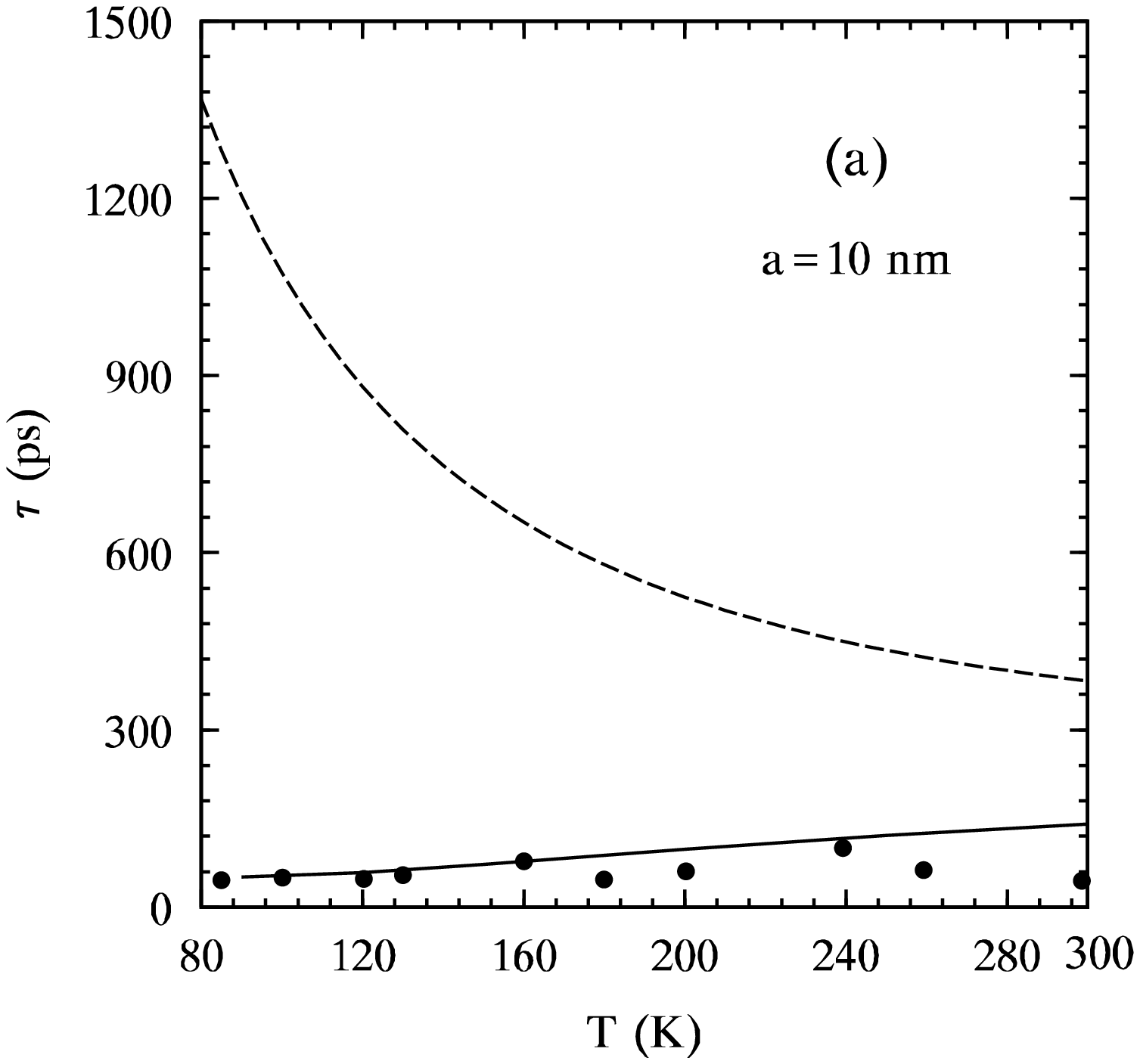,width=9.5cm,height=8cm,angle=0}
  \psfig{file=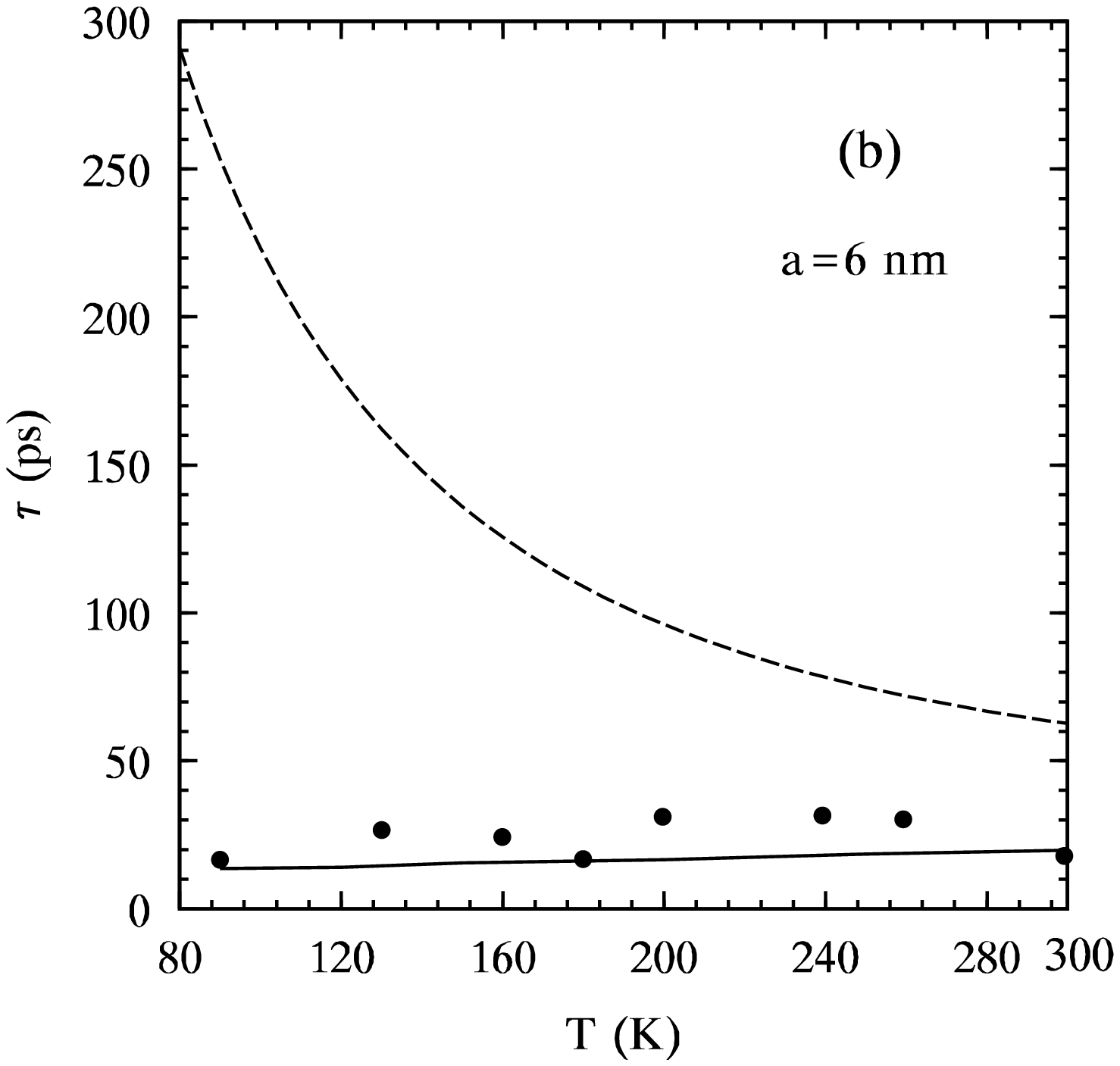,width=9.5cm,height=8cm,angle=0}
  \caption{The spin dephasing time as a function of the temperature
  for two GaAs quantum wells with different well width:
  (a) $a$=10~nm; (b) $a$=6~nm.}
  \label{fig1}
\end{figure}

In Fig.\ 1 the spin dephasing time is plotted as a function of the
temperature for 10\ nm QW (a) and 6\ nm QW (b). The corresponding
experimental data are plotted as dots.  For comparison, the spin
dephasing time calculated from the single particle theory Eq.\
(\ref{eq:taus_ani}) from the {\em same} parameters is also given as dashed
curve in the same figure.  It is seen exactly from the figure that our
results fit very well with the experiment data.  In contrast, the spin
dephasing time predicted by the single particle model is about one
order of magnitude {\em slower} than the experiment data. Moreover, it is
noticed that when the temperature increases, both our theory and
experiment data indicate that the spin dephasing time slightly
{\em increases} with the temperature whereas the single particle model
predicates a dramatic {\em decrease} of the spin dephasing time with the
increase of the temperature.  

These results clearly demonstrate the important role the many-body
effect plays in the spin dephasing. The single particle theory does
not count for the many-body dephasing due to the
inhomogeneous broadening inherited in the DP term.\cite{wu_pss_2000,%
wu_ssc_2002,wu_jpsj_2001,wu_epjb_2000,wu_js_2001} 
Therefore, it is easy to understand why the single particle
model gives much slower spin dephasing time.

Taking the inhomogeneous broadening effect into account, one can also
understand the temperature dependence of the spin dephasing time: as
the temperature increases, the inhomogeneous broadening is reduced as
the electrons are distributed to the wider ${\bf k}$-states. As the result,
the number of electron occupation on each ${\bf k}$ state around the
Fermi surface is reduced. Consequently the inhomogeneous broadening
and therefore the many-body spin dephasing are suppressed. 
It is further noted that this reduction is mild as a function
of the temperature. Therefore, the temperature dependence is quite
mild.
 
In summary, we have shown evidence of the many-body effect in the spin
dephasing for $n$-typed narrow QW's where the spin dephasing is
dominated by the many-body effect.  It is noted that the small well
width not only limits electrons to the lowest subband, but also
enhances the DP term as $\langle k_z^2\rangle=(\pi/a)^2$ in Eq.\
(\ref{eq:dp}) is greatly enhanced for small width and it is much
larger than $k_F^2$ for the density around $10^{11}$\ cm$^{-2}$.
Consequently the inhomogeneous broadening is also greatly enhanced.
For wide QW or bulk sample, the DP term is not enhanced. We notice
from the experiment that the temperature dependence of the spin
dephasing time is similar to what predicted by the
single particle 
theory, {\em ie.}, the spin dephasing time decreases with the
temperature.  A full many-body quantum kinetic calculation for the
quantatative accurate spin dephasing time is very time consuming and
is still missing. Our earlier qualitive calculation for bulk GaAs has
shown the decrease of the spin dephasing time with the
temperature.\cite{wu_pss_2000} A thorough investigation is still
underway to find out how large the many-body effect is for those
samples. 

\bigskip
MWW is supported by the  ``100 Person Project'' of Chinese Academy of
Sciences and Natural Science Foundation of China.


\end{document}